\documentclass[11pt]{cernrep}
\usepackage{graphicx}
\usepackage{here}

\begin{document}

\title{A new hard QCD phenomenon - proton diffraction into three jets}

\author{L.~Frankfurt$^1$,  M.~Strikman$^2$}
\institute{Nuclear Physics Dept.,
        Tel Aviv University, Israel$^1$,
Department of Physics, PSU, USA$^2$}
        
\maketitle
\begin{abstract}
We argue that the process of the proton diffraction into three jets
in proton-nucleus scattering at LHC will allow 
to investigate the light-cone wave function of protons,
generalized gluon densities in nuclei at small x, as well as the 
interplay of the color transparency and opacity phenomena. 
\end{abstract}

\section{Introduction}

During the last ten years a number of new hard small x phenomena have been 
observed and calculated in QCD based on the  QCD factorization theorems for 
inclusive DIS and hard exclusive processes. This includes (i) the observation 
at HERA of a fast increase  of parton distributions with energy which 
is reasonably described within the QCD evolution equation approximation.
\footnote{A comprehensive review of the HERA data on inclusive and exclusive 
processes and the relevant theory is given  in  \cite{AC}.} 
(ii)the discovery of the color transparency phenomenon in the pion coherent 
dissociation into two high $p_t$ jets \cite{ashery} which is consistent 
with the predictions of \cite{FMS93,FMS2000}, 
(iii) the observation of various regularities in the exclusive 
electroproduction of vector meson at HERA induced by longitudinally 
polarized photons and photoproduction of mesons with hidden heavy flavor
which is consistent with the predictions of \cite{Brod94}. These processes 
in a wide kinematic range of small x
have provided effective ways of studying the interaction of small 
colorless dipoles with hadrons at high energies and of studying the 
hadron wave functions  in the minimal 
Fock space configurations. Here we outline new QCD phenomena \cite{FELIX}
which may become important at LHC because their cross sections are rapidly 
increasing with energy.

Fast increase with energy of cross sections of hard diffractive processes 
predicted within the leading twist (LT) approximation would be in variance 
with the unitarity of $S$-matrix for the interaction with a hadron target of 
spatially small wave package of quarks and gluons at given impact parameter. 
This new phenomenon may reveal itself at sufficiently small x and/or 
sufficiently heavy nuclei. In the structure functions off a proton this 
physics is masked by the significant contribution for the hard collisions
off nucleon periphery. Thus the unitarity of the S matrix  does not preclude 
fast increase of structure functions of a nucleon as fast as  
$\propto {\ln (1/x)}^{3}$ \cite{fmgs}. Hence no dramatic signals for the 
breakup of DGLAP regime are expected in this case.
On the contrary, a breakdown of the LT 
approximation in the hard diffractive processes off nuclei leads to the 
drastic  changes of the dependence of cross sections on atomic number,
the incident energy and the  transverse momenta of the jets.

\section{Diffraction into three jets}
\label{3jets}

A nucleon (meson) has a significant amplitude to be in a configuration where 
valence partons are localized in a small transverse area together with the 
rest of the partons. These configurations are usually referred to as 
{\it minimal} Fock space configurations -  $\left|3q \right>$ . The 
significant amplitude of the decay $\pi\rightarrow leptons$ as well as the 
observed significant cross  section of the process: $\pi+A\rightarrow 2jet +A$ 
\cite{ashery} is the experimental evidence for the significant probability 
of such configurations in mesons. In the case of proton the amplitude of the 
three quark configuration can be estimated within QCD inspired models of 
proton and in the long run it can be calculated in the lattice QCD. 
Note that the knowledge of this amplitude is important for the unambiguous 
calculation of proton decay within the Grand Unification models.

Hadrons, in small size configurations of the transverse size $d$, interact 
with a small cross section $\propto d^2 xG_N(x,Q^2)$, where $Q^2\sim 10/d^2$.
The factor $xG_N(x,Q^2)$ leads to a fast increase of the {\it small }
cross section with energy. In the case of the "elastic
scattering" of such a proton configuration off another proton, this
three-quark system with large relative momenta should preferentially
dissociate diffractively into a system of three jets with large transverse
momenta $p_{ti} \sim \pi~b^{-1}$, where $b $ is the transverse size of the
minimal configuration. The kinematics of this process is presented at the lego
plot of Fig.1 for the case of the coherent scattering off nuclei(proton).

\begin{figure}
\begin{center}
\includegraphics[width=10.5cm]
{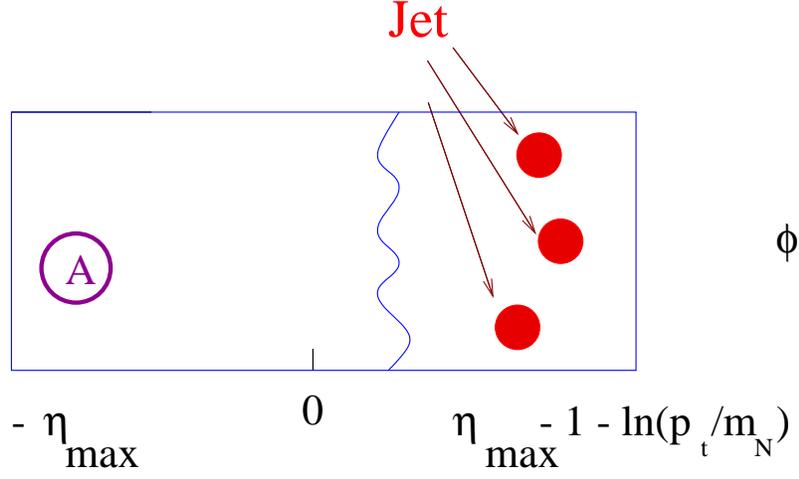}
\caption{LEGO plot for diffraction of proton into three jets
}
 \end{center}
\label{fig:nuc}
\end{figure}

The production cross-section for the three jets can be evaluated based on the 
kind of QCD factorization theorem deduced in \cite{FMS2000}. The 
cross-section is proportional to the square of the gluon density in the 
nucleon at $x\approx {M^2_{3 jets}/ s}$, and virtuality
$\sim (1-2) p_{t}^2$\cite{FS98}. The distribution over the fractions of 
the longitudinal momentum carried by the jets is proportional to the
square of the light 
cone wave function of the $\left |3q \right>$ configuration,
$\psi(z_1, z_2, z_3, p_{t1}, p_{t2}, p_{t3})$ and large transverse
 momenta where
\begin{equation}
\psi(z_1, z_2, z_3, p_{t1}, p_{t2}, p_{t3})\propto z_1z_2z_3 
\sum_{i\neq j} {1\over p_{ti}^2}{1\over p_{tj}^2}
\end{equation}

It is implied that proper renormalization procedure is performed which 
accounts for cancellation of infrared divergences. Or the light cone 
gauge $p_{N,\mu}A_{\mu}=0$ should be used to describe diffractive
fragmentation region: where there are no infrared divergences. Here 
$p_{N}$ is the momentum of diffracting proton. Besides summation over 
collinear radiation is included in the  definition of the jets.

Application of the
QCD factorization theorem resolves puzzle with the fact that 
a hadron contains infinite 
number of partons - other partons are hidden into structure functions of a 
target. Hence the diffraction of a proton into three jets provides important 
information about the short-distance quark structure of the proton, and also 
provides unique information about the longitudinal momentum distribution in 
the $\left |3q \right>$ configuration at high $p_t$. 
The amplitude of the process can be written as
\begin{equation}
A=\left[-{\vec \Delta}
\psi(z_1, z_2, z_3, p_{t1}, p_{t2}, p_{t3})\right]\left[{2\pi^2\over 3}
 \alpha_s(xG(x,Q^2,t)_{skewed})\right].
\end{equation} 
Here  $\Delta=\sum^3_{i=1}{\partial^2\over\partial p_i^2 }$. An additional 
 factor of 2
 in the numerator  as compared to the pion case is due to the possibility for
two gluons to be attached to different pairs of quarks.
The $p_t$ and longitudinal momentum 
fraction, $z_i$ dependence of the cross section is given by 
\begin{eqnarray}  
\frac{d\sigma}{dz_{1} dz_{2} dz_{3} 
d^{2}p_{t1}  d^{2}p_{t2} d^{2}p_{t3}}=
 c_N |\alpha_{s} xG(x,Q^{2})|^{2}
\frac{|\phi_{N}(z_{1},z_{2},z_{3})|^{2}}{|p_{t1}|^{4}
|p_{t2}|^{4} |p_{t3}|^{4}}F^2_{2g}(t) \nonumber \\ 
\delta(\sum p_{ti}-\sqrt{-t}) 
\delta(\sum z_{i}-1).
\end{eqnarray}
The coefficient $c_N$  is calculable in QCD, and function 
$\phi_{N}(z_{1},z_{2},z_{3})=z_1z_2z_3$. 
The two-gluon form factor of a nucleon is known to some extent from 
the hard diffractive processes \cite{FS2002}.
The numerical estimate
of the cross section 
of production of three jets with $p_t$ of 
one 
of the 
jets larger than a given one and integrated over all other variables gives 
for $p_t\ge$ 10 GeV/c:
\begin{equation}
\sigma(pp\to 3jets + p)\propto \frac{|\alpha_s xG(x,Q^2)|^{2}} 
{p_t^{8}} \propto 10^{-6\div -7} \left(\frac{10 GeV}{p_{t}}\right)^{8}mb
\end{equation}
for  LHC energies.  A pressing question is whether this cross section 
will grow up to the LHC energies as it is assumed in this estimate
based on extrapolations of $G_N(x,Q^2)$ to $x \sim 10^{-5}$.
The probability of the $\left |3q \right>$ configuration is estimated
using a phenomenological fit to the probability of configurations of
different interaction strengths in a nucleon  (cf.  \cite{FS98,FMS94}).

A study of the same process in the 
proton nucleus collisions would provide an 
unambiguous test of the dominance of hard physics in this process. The cross 
section of this process should grow with $A$ as 
${A^{2}\over R_A^2} {G_A^2(x,Q^2)/A^2 G_N^2(x,Q^2)}$. Account of skewedness 
of gluon distribution will lead to some enhancement of absolute value of 
cross section but will not influence strongly A-dependence. 

Our estimates indicate for the relevant LHC kinematics - 
$x \sim 10^{-5}, Q^2_{eff}\sim p_T^2$ 
nuclear shadowing in the gluon channel is a rather small correction reducing 
the A-dependence of cross section by a factor $A^{0.1}$. The A-dependence of 
total inelastic cross section is $\propto A^{0.7}$. Hence we expect 
that counting rate per one inelastic interaction will be enhanced in $pA$ 
collisions for heavy nuclear target by a factor $\propto A^{0.55}$.
At the same time the background from the soft diffraction should
be strongly reduced. Indeed, if we estimate the A-dependence of the soft 
diffraction based on the picture of color fluctuations which provides  
a good description of the total cross section of the inelastic
diffraction off nuclei at fixed target energies \cite{FMS93}
we find $\sigma_{pA}^{inel.~diff.} \propto A^{0.25}$.
Note also that scattering off nuclei has an obvious advantage in terms of 
selecting coherent processes since practically all inelastic interactions 
with nuclei would lead to the emission of neutrons at the zero angle.

A competing process is the proton diffraction into 3 jets off the 
nuclear Coulomb field. It follows from the theorem proved in \cite{FMS2000}
that in the LO over $\alpha_s$ the dominant contribution is given in the 
Weizecker-Williams representation by the  transverse photon interaction 
with the external quark lines. The amplitude of the process is 
\begin{equation}
{-e^2\psi_N(z_i,p_{ti})\over q^2-t_{min}}Z F_A(t){4s \over M^2(3jet)}
\sum_i{e_i(\vec p_{ti}\cdot \vec{q_t})\over z_i},
\end{equation}
where $e_i$ is the electric charge of the quark in the units of the electric
 charge of the electron, $q^2$ is the virtuality of the exchanged 
photon. 

In the kinematics where the LT  dominates in the QCD mechanism
for production
of three jets the e.m. process is a small correction.
However if the screening effects (approach to the black body limit) occurs
for the discussed $p_t$ range the e.m. process may compete with the 
QCD contribution.
At the same time in the case of the 
nucleus break up the e.m. process has no singularity at $q^2=0$ besides
the QCD process is $\propto A$ and much larger than the
e.m. process. Note also that in the case of QCD interactions 
one can consider the processes with large rapidity gap.
They have the same $z_i,p_{ti}$ dependence as the coherent process but have
significantly larger cross section.

Another interesting group of hard processes is proton diffraction into two
high $p_T$ jets and one collinear jet. These processes are in general 
dominated at high energies by collisions of two nucleons at large impact 
parameters (otherwise multiple soft interactions would destroy the coherence).
Hence  the A-dependence of the dominant term 
should be similar to that for $\sigma_{pA}^{inel.~diff.}$. However already
fluctuations with cross section of $\sim 40 mb$ would have much  stronger
A-dependence $\sim A^{0.7}$. Thus the study of the A-dependence of coherent 
diffraction will allow to filter out the smaller than average components 
in the nucleons at the LHC energies.

\section{Violation of LT approximation and new QCD strong interactions}.

Energy dependence of cross section in the LT
 approximation follows from the
properties of gluon distribution within the target.  
Hence for the virtualities probed in the three jet production
the energy dependence is 
$\propto  x^2G^2(x,Q^2_{eff}) \propto s^{0.8}$
for $Q^2_{eff}\sim 100 -1000 GeV^2$, and $10^{-5}\le x \le 10^{-2}$.
In the black body limit the inelastic diffractive processes are strongly 
suppressed and hence the energy dependence of cross section the three
 jet production
integrated over $p_T\ge p_{T,0}$  should slow down as only scattering off the 
grey edge of nucleus (nucleon)  will effectively survive. 
For the $Q^2$ relevant for this process the black body
limit maybe an interesting hypothesis for the scattering 
off heavy nuclei only.
It would manifest itself in a strong decrease of the $A$-dependence of
the three jet production at $p_t$ corresponding to the black body limit.
At sufficiently small $x$ for a heavy  nuclear target:
${d\sigma(pA\rightarrow 3jet +A)\over dp_t^2} \propto A^{1/3}$.  
This is in a striking contrast with the expectations for the A
dependence parton distributions : $\propto A^{2\over 3}$.

We thank  GIF and DOE for support.

\end{document}